\documentclass[12pt, hyper]{article}
\usepackage{amsmath,amsopn,amssymb,amsfonts}
\usepackage{amsthm}
\usepackage{epsfig}
\usepackage{verbatim}
\usepackage{multirow}
\usepackage{arydshln}
\usepackage{amsthm}
\usepackage[margin=1cm]{caption}

\def\CE{{\cal E}}
\def\CH{{\cal H}}
\def\CL{{\cal L}}
\def\CN{{\cal N}}
\def\CQ{{\cal Q}}

\def\a{\alpha}

\def\z{\zeta}

\def\m{\mu}
\def\r{\rho}
\def\t{\tau}

\def\S{\Sigma}

\DeclareMathOperator*{\Res}{Res}
\DeclareMathOperator*{\JKres}{JK-Res}

\usepackage{cleveref}
\crefname{section}{¤}{¤¤}
\Crefname{section}{¤}{¤¤}

\begin{document}

\begin{titlepage}
\vfill
\begin{flushright}
{\tt \normalsize EFI-14-20}\\

\end{flushright}
\vfill
\begin{center}
{\Large\bf Comments on Quantum Higgs Vacua}

\vskip 1cm

Jeffrey A. Harvey,
David Kutasov and
Sungjay Lee

\vskip 5mm
{\it Enrico Fermi Institute and Department of Physics
University of Chicago \\
5620 Ellis Av., Chicago Illinois 60637, USA}

\end{center}
\vfill

\begin{abstract}
\noindent
We study two-dimensional $\CN=(4,4)$ supersymmetric gauge theories
which do not have classical  Higgs branches, but are believed to have
isolated quantum Higgs vacua. We provide arguments for the existence of
such vacua based on brane dynamics in string theory and the supersymmetric partition sums
of the gauge theories.
\end{abstract}

\vfill
\end{titlepage}

\renewcommand{\thefootnote}{\#\arabic{footnote}}
\setcounter{footnote}{0}

\section{Introduction}

Many two dimensional supersymmetric gauge theories have a classical moduli
space of vacua consisting of Higgs and Coulomb branches connected at the
origin.\footnote{In general there are also mixed branches;
our discussion is easy to generalize to include them.}
Due to the large infrared fluctuations associated with scalar fields in two dimensions,
one might expect the low energy theory to be a sigma model
on this moduli space. Instead, it often splits into a direct sum of separate theories
associated with the Coulomb and Higgs branches \cite{Witten:1997yu}.

A simple example of this phenomenon occurs in $\CN=(4,4)$ supersymmetric QED with $F$ charged
hyper multiplets.
Classically, this theory has a one (quaternionic) dimensional Coulomb branch
and an $F-1$ dimensional Higgs branch, that intersect at the origin of both. Quantum mechanically,
it splits in the infrared into two decoupled SCFT's, which correspond to the two classical branches
and in general have different central charges.

From the perspective of the Coulomb branch theory the classical origin is pushed
to infinity by a quantum correction to the metric. More precisely,
the one-loop exact metric on the Coulomb branch is \cite{Douglas:1997vu}
\begin{align}
  ds^2 = \left( \frac{1}{g^2} + \frac{F}{|\vec \phi|^2} \right) d\vec \phi \cdot d\vec \phi\ ,
\end{align}
where $\vec \phi$ are the scalars in the $(4,4)$ vector multiplet, and $g^2$ is the gauge coupling
constant. The origin of the Coulomb branch, $\vec\phi=0$, which classically is a finite point, lies
quantum mechanically at the bottom of an infinite throat.
This is regarded as a signal of the above decoupling in this branch \cite{Witten:1997yu}.
It is also useful to recall for future reference that the
Coulomb branch can be lifted by turning on an FI term $\xi$ for the $U(1)$ factor
in the gauge group, which breaks the gauge symmetry and leaves only the Higgs branch theory.

An interesting special case of the above theory is $F=1$, i.e., one charged hyper multiplet.
For $\xi\not=0$ this theory has an isolated Higgs vacuum in which the gauge symmetry is broken.
For $\xi=0$, the classical analysis suggests that there is a Coulomb branch, but no Higgs vacuum.
However, there are reasons to believe that there is actually a separate isolated
Higgs vacuum which does not contain any light excitations and owes its existence to quantum dynamics.
This vacuum is important for describing fivebranes in  matrix string theory,
and for understanding the D$1$-D$5$ system.

A similar quantum Higgs vacuum is expected to appear in theories with higher rank gauge groups,
such as $\CN=(4,4)$ supersymmetric QCD with gauge group $U(N)$ and $F=N$ hyper multiplets
in the fundamental representation. One can think of its existence as a
consequence of the duality satisfied by the Higgs branch theory, under which $(N,F)\to (F-N,F)$.

In this note we discuss these vacua from two points of view: brane dynamics in string theory, and the supersymmetric
index of the relevant gauge theory. In section 2 we review the embedding
of $\CN=(4,4)$ SQCD in string theory as the low energy theory on a system of intersecting NS$5$-branes and D-branes,
which makes the quantum vacuum structure more transparent, partly because it makes the above Seiberg-like duality manifest.
We use this description to understand the isolated quantum Higgs vacua,
and discuss briefly the closely related problem of the behavior of the duality under mass deformations.
In section 3 we review the definition of the supersymmetric index of  $\CN=(0,2)$ supersymmetric theories
computed recently in \cite{Benini:2013nda,Benini:2013xpa} using localization (see also \cite{Gadde:2013dda}).
We then explain how the supersymmetric
index of these gauge theories can be used to deduce the existence of the isolated quantum vacua.

\section{Brane Perspective}

\begin{figure}[h]
\begin{center}
  \includegraphics[width=15cm]{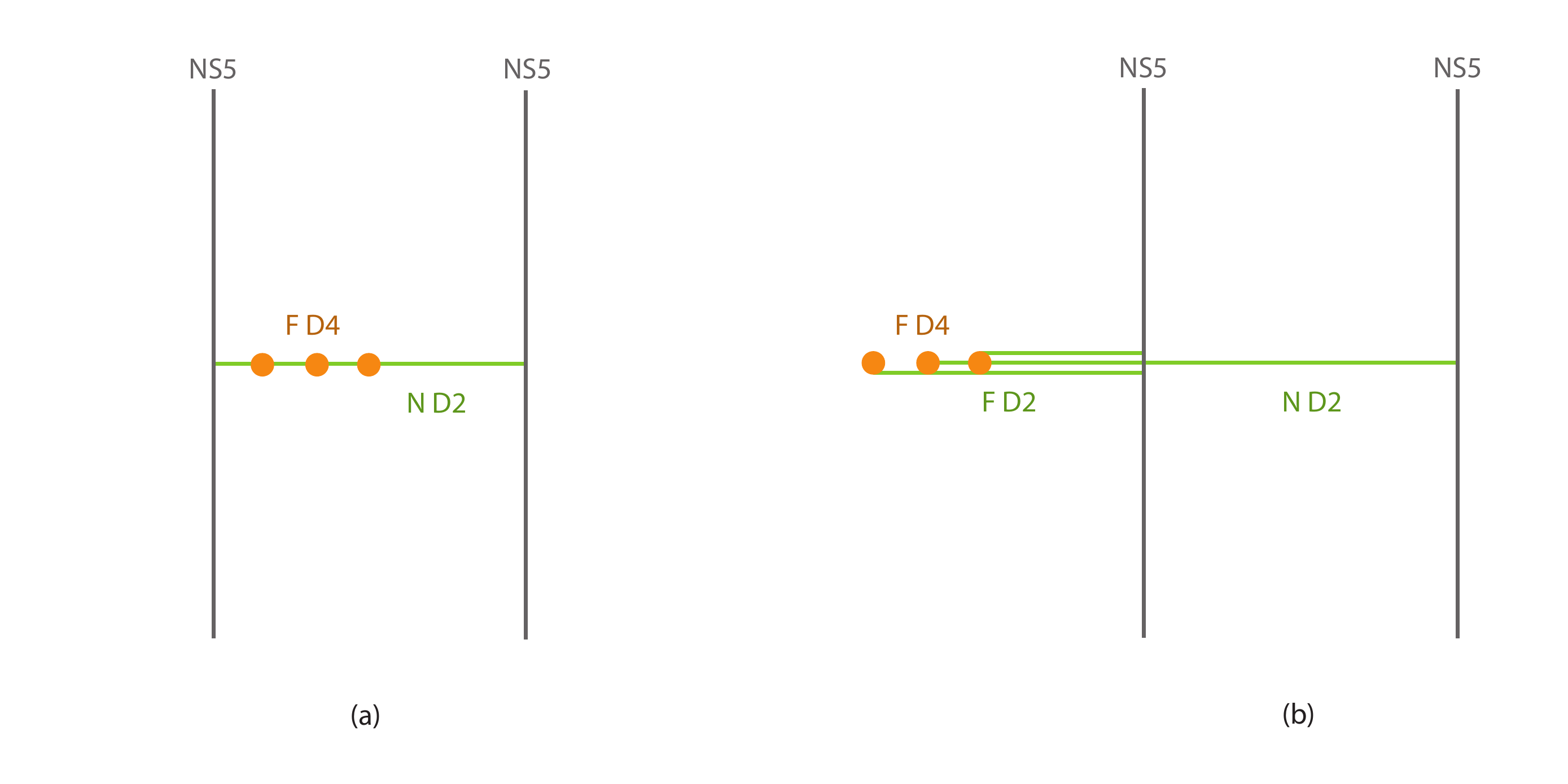}
  \caption{IIA brane realizations of $\CN=(4,4)$ SQCD.}\label{figure1}
\end{center}
\end{figure}

Systems of NS$5$-branes and D-branes in string theory are useful for studying the low
energy behavior of many supersymmetric and non-supersymmetric vacua of SYM theories
in various dimensions (see e.g. \cite{Giveon:1998sr} for a review), and this is the case here as well.
The brane system relevant for describing $\CN=(4,4)$
SQCD\footnote{A dimensional reduction of that of \cite{Hanany:1996ie},
which was used to study $\CN=4$ supersymmetric gauge theories in $2+1$ dimensions.}
is depicted in figure \ref{figure1}(a).
The $\CN=(4,4)$ $U(N)$ gauge multiplet lives on $N$ D$2$-branes stretched in the $x^6$ direction
between two NS$5$-branes whose world volume fills the $(2345)$ directions.
Both types of branes are extended in the $1+1$ dimensional spacetime labeled by $(01)$.
The hypermultiplets live at
intersections of the D$2$-branes and $F$ D$4$-branes stretched in $(01789)$.
To summarize, the worldvolume directions of the various branes are:
\begin{center}
\setlength{\tabcolsep}{0.25cm}
\begin{tabular}{c|cc|cccc|c|ccc}
  \hline
  & 0 & 1 & 2 & 3 & 4 & 5 & 6 & 7 & 8 & 9  \\
  \hline \hline
  NS5 & $\times$ & $\times$ & $\times$ & $\times$ & $\times$ & $\times$ & & & &  \\
  D2 & $\times$ & $\times$ &  &  &  &  & $\times$ & & &  \\
  D4 & $\times$ & $\times$ & & & & & & $\times$ & $\times$ & $\times$ \\
  \hline
\end{tabular}\
\end{center}

To study the isolated quantum Higgs vacua it is convenient to use an alternative representation of this system,
depicted in figure \ref{figure1}(b). The representations of figures  \ref{figure1}(a) and (b) are related by the Hanany-Witten transition
\cite{Giveon:1998sr,Hanany:1996ie}. Consider, for instance, the case $N=1$ (gauge group $U(1)$).
The classical Coulomb branch is parametrized by the location  in the $(2345)$ directions of the color D$2$-brane
stretched between the fivebranes. The Higgs branch is obtained in figure \ref{figure1}(a) by letting
the D$2$-brane split into segments on the D4-branes. The $F-1$ moduli correspond to the positions of the segments stretched between adjacent
D4-branes in the $(789)$ directions.\footnote{The fourth modulus in the multiplet can be identified with the $6$ component of the gauge field on the D$2$-branes.}
In figure \ref{figure1}(b)
the Higgs branch is described by reconnecting the color D$2$-brane to the longest flavor D$2$-brane,
and allowing the segments of the resulting two brane connecting adjacent D$4$-branes to move in the
$(789)$ directions \cite{Giveon:1998sr}.
The two branches intersect at the origin of both, which is depicted in figure \ref{figure1}.

Turning on an FI term $\xi$ for the $U(1)$ corresponds in the brane description
to a relative displacement of the two NS$5$-branes in the transverse directions $(789)$.
As is clear from figure \ref{figure1}, this lifts the classical Coulomb branch,
and leaves behind only the Higgs branch.

For $N=F=1$ we recover the structure mentioned in section 1.
For $\xi\not=0$ we find a vacuum, which from the brane perspective is
described by a configuration in which the color and flavor D$2$-branes in figure \ref{figure1}(b)
combine into a single D$2$-brane that connects the right NS$5$-brane to the D$4$-brane,
never intersecting the second NS$5$-brane. Since the low energy theory on such a D$2$-brane is massive \cite{Hanany:1996ie},
this vacuum is isolated.

For $\xi=0$, the classical picture of figure \ref{figure1} suggests that the
above isolated state becomes identical to the origin of the Coulomb branch,
and in particular there is no independent isolated vacuum.
However, as we mentioned, quantum mechanically we expect such a vacuum to exist.
To understand this from the brane point of view we recall that
the classical configurations of figure \ref{figure1} are corrected at finite
$g_s$ by brane bending effects \cite{Callan:1997kz,Witten:1997sc}. A D$2$-brane ending on an NS$5$-brane
can be thought of as a charged particle in the five dimensional gauge theory
obtained by reducing the fivebrane theory on $x^1$ (or, equivalently, a string charged under the self-dual $B_{\mu\nu}$
field in the six dimensional IIA fivebrane theory).
Thus, the gauge field on the fivebrane has a non-trivial profile $F_{0r}\sim 1/r^3$,
where $r$ is the radial direction in $(2345)$.
Since this configuration is BPS, there is also a non-trivial profile of the
scalar field on the fivebrane $x^6(r)\sim 1/r^2$.

\begin{figure}[t]
\begin{center}
  \includegraphics[width=15cm]{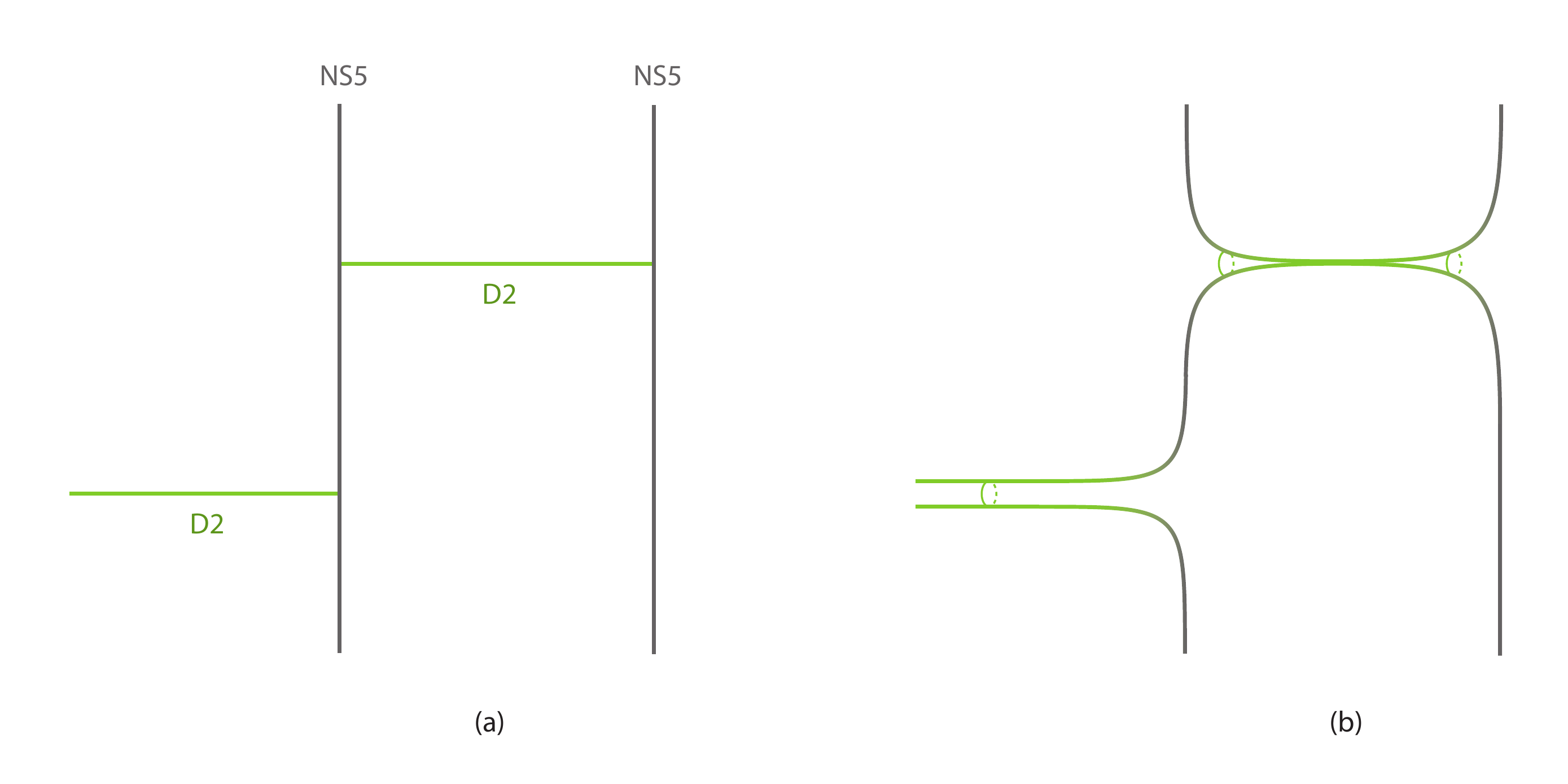}
  \caption{Classical (a) and quantum (b) descriptions of the Coulomb branch of $U(1)$ $\CN=(4,4)$
  SQED with one charged hyper multiplet.}\label{figure2}
\end{center}
\end{figure}
\begin{figure}[h]
\begin{center}
  \includegraphics[width=15cm]{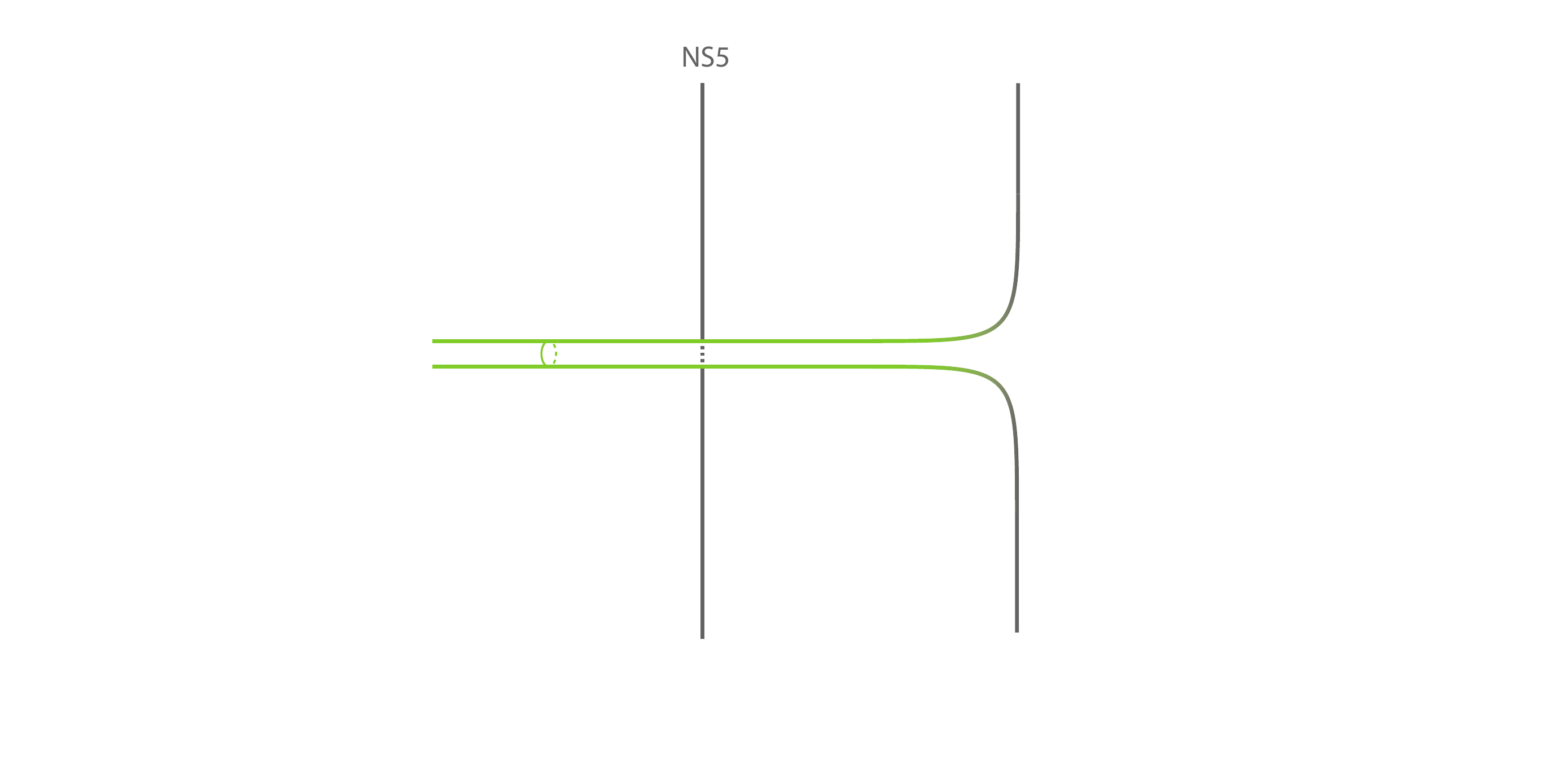}
  \caption{The quantum Higgs vacuum of the theory of figure \ref{figure2}.}\label{figure3}
\end{center}
\end{figure}

This has an important effect on the vacuum structure.
A generic point on the Coulomb branch, that classically
is described by the configuration of figure \ref{figure2}(a),
is replaced at finite $g_s$ by that of figure \ref{figure2}(b).
In particular, the kinetic term for the position of the color $D2$-brane in
the $(2345)$ directions, $\vec \phi(x^\mu)$, which classically is canonical,
$\CL_\text{kin}\sim |\partial_\mu \vec \phi|^2$, develops a singularity at $\vec \phi=0$,
$\CL_\text{kin}\sim |\partial_\mu \vec \phi|^2/|\vec \phi|^2$. Thus, the origin of the
Coulomb branch is pushed to infinity, in agreement with the discussion of \cite{Witten:1997yu,Douglas:1997vu}.
The supersymmetric brane configuration depicted in figure \ref{figure3} represents the isolated quantum Higgs vacuum of the theory.

The above discussion easily generalizes to the case of $U(N)$ gauge theory
with $F=N$ hypermultiplets (see figure \ref{figure1}). It is obvious in the brane
description that the classical theory (with $\xi=0$) has an $N$ dimensional Coulomb branch,
parametrized by the positions of the $N$ color D$2$-branes, and no vacua with
lower rank gauge group. Quantum mechanically, the vacuum structure is much richer:
each of the $N$ color D$2$-branes can combine with a flavor D$2$-brane to form the configuration
of figure \ref{figure3}, thereby reducing the rank of the gauge group by one unit.
When all $N$ color D$2$-branes do this, one finds an
isolated vacuum with no light excitations.

One can take a broader viewpoint and consider the theory with general $N$ and $F$ in figure \ref{figure1}.
This theory has a rich vacuum structure that depends on $N$, $F$. For $F<N$, a Higgs branch does not exist,
and if we turn on a non-zero FI parameter $\xi$ for the $U(1)$, there is no supersymmetric vacuum.
For $F>N$ there is a Higgs branch of the classical moduli space in which the gauge symmetry is fully broken.
Its dimension is given by $N(F-N)$. This branch gives rise in the quantum theory to a CFT
with central charge  $c=6 N(F-N)$.

This CFT enjoys a duality $N\to F-N$, which is due to the fact that the Higgs branches of the $U(N)$
and $U(F-N)$ theories with $F$ flavors coincide. In the gauge theory this is familiar from studies
of $\CN=2$ SQCD in four dimensions \cite{Antoniadis:1996ra}. In two dimensions it was demonstrated
that the two-sphere partition functions \cite{Benini:2012ui,Doroud:2012xw} and elliptic genera \cite{Benini:2013xpa} of
these two gauge theories are the same, which supports the duality.
In the brane language it is a simple generalization of the results of \cite{Elitzur:1997fh}.

For $F=N$ the electric theory is the Higgs branch of the $U(N)$ theory with $N$ flavors,
while the magnetic theory becomes a $U(0)$ theory, i.e. a supersymmetric vacuum with no light excitations.
This supersymmetric vacuum is the isolated Higgs vacuum we discussed above.

We note in passing that the set of isolated vacua becomes richer when we consider deformations.
For example, consider turning on (equal) masses for $F_1$ of the $F$ fundamental hypermultiplets.
\begin{figure}[t]
\begin{center}
  \includegraphics[width=15cm]{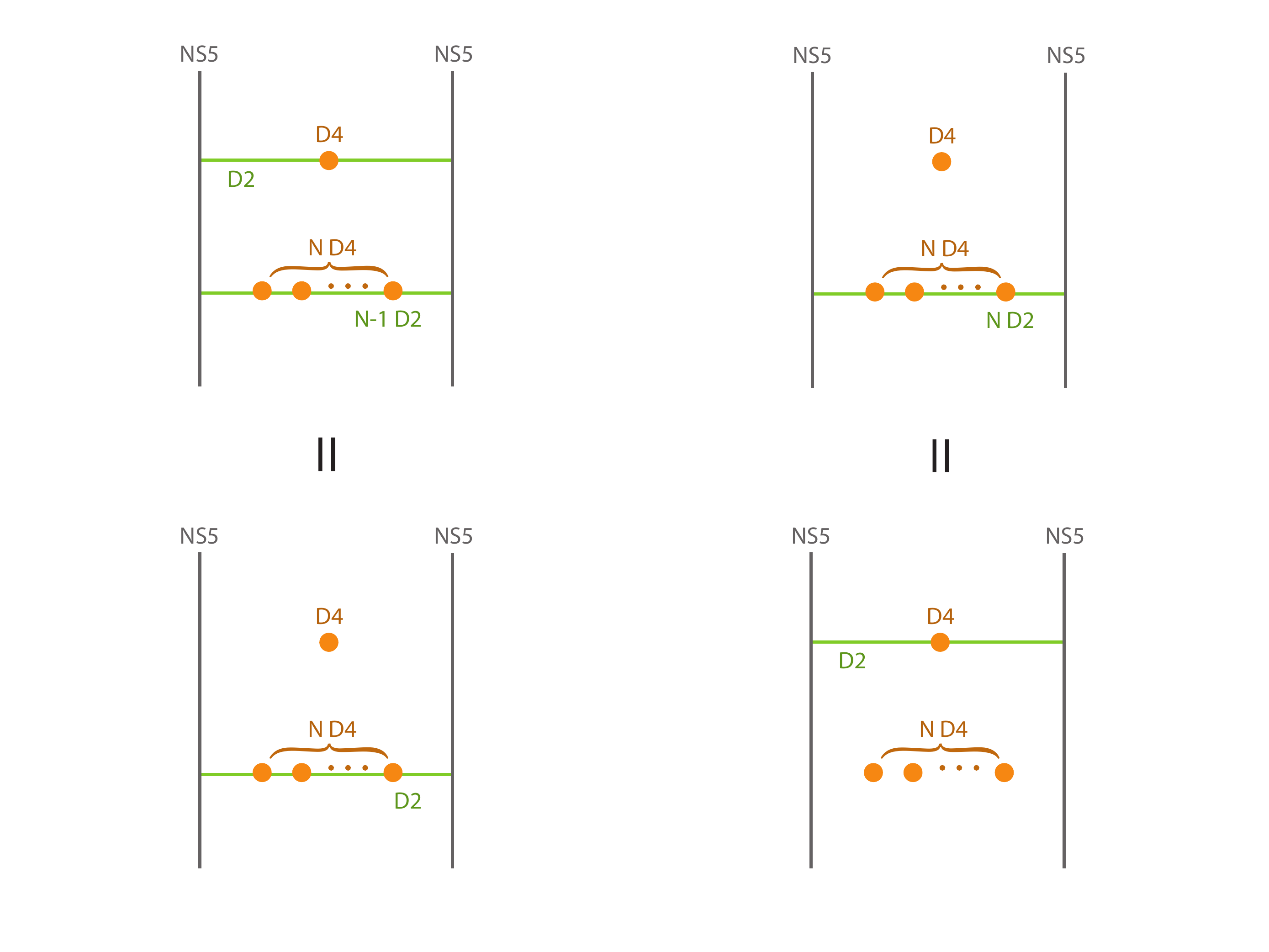}
  \caption{Map of Higgs vacua of a $U(N)$ gauge theory with $F=N+1$ flavors (top) and the dual $U(1)$ theory with $N+1$ flavors
 (bottom) when one of the flavors has a non-zero mass.}\label{figure4}
\end{center}
\end{figure}
In the brane language this corresponds to displacing $F_1$ of the D$4$-branes from the origin of the $(2345)$
space to a particular point while leaving the remaining $F_2=F-F_1$ D4-branes at the origin.
In order for the theory to have a Higgs branch, we need to divide the D$2$-branes
into two groups, of $N_1$ and $N_2$ with $N_1+N_2=N$, and displace them so that
they intersect the corresponding sets of D$4$-branes;
see figure \ref{figure4} for a simple example.
The existence of a Higgs vacuum implies that $0\le N_i\le F_i$, which can be equivalently written as
\begin{align}
  \max(F_1-\hat N,0)\le N_1\le \min(F_1,N)
\end{align}
where $\hat N=F-N$. It is easy to see that this condition is invariant under the duality map
$N\to \hat N$, $N_i\to \hat N_i=F_i-N_i$, in agreement with the fact that the low energy theory
reduces in this case to a direct product of CFT's at the two intersections of $N_i$ and $F_i$ branes,
with central charges $c_i=6N_i(F_i-N_i)$.

Whenever one of the $N_i$ coincides with the corresponding $F_i$,
we get an isolated vacuum in that theory, which corresponds under the duality map
to a theory with no light excitations but a supersymmetric vacuum.
Thus, the existence of the quantum Higgs vacua is necessary
for the consistency of the $(4,4)$ duality with deformations.

\section{Supersymmetric Index}

Two dimensional $\CN=(4,4)$ $U(N)$ SQCD coupled to $F$ hypermultiplets
in the fundamental representation has a global $SU(F)$ flavor symmetry, as well as an $SU(2)_1 \times SU(2)_2 \times SU(2)_3$ R-symmetry,
under which the vector multiplet scalars transform as $({\bf 1},{\bf 2},{\bf 2})$ and
the hyper multiplet scalars transform as $({\bf 2},{\bf 1},{\bf 1})$.
The four right-moving supercharges $\CQ_+^{a\a}$  transform as $({\bf 2},{\bf 2},{\bf 1})$,
while the four left-moving supercharges $\CQ_-^{a\dot \a}$ transform as
$({\bf 2},{\bf 1},{\bf 2})$. The indices $\a,\dot\a$ correspond to doublets under $SU(2)_2$ and
$SU(2)_3$, respectively.  For $F>N$, when the Higgs branch theory is non-trivial,  $SU(2)_2$ and $SU(2)_3$ become parts of the infrared right
and left-moving $\CN=4$ superconformal algebras, respectively.

Choosing particular supercharges $\CQ_+^{++}$ and $\CQ_+^{--}$ that generate
an $\CN=(0,2)$ subalgebra of the full $\CN=(4,4)$ SUSY algebra
(i.e. $\{\CQ_+^{++}, \CQ_+^{--}\} = \bar L_0$), we can define a supersymmetric partition sum as follows
\begin{align}
  \CE(\t;z,\z_i) = \text{Tr}_{\CH_\text{RR}} \Big[ (-1)^F q^{L_0} \bar q^{\bar L_0}
  e^{-2\pi i z J_R}  e^{-2\pi i \z_1 (J_1-J_2)} e^{-2\pi i \z_2 \cdot J_f }
   \Big]\ ,
  \label{def}
\end{align}
where $\{J_1,J_2,J_R,J_f\}$ are Cartan subalgebra generators of the R-symmetry group
$SU(2)_1\times SU(2)_2\times SU(2)_3$ and the flavor group $SU(F)$.
Note that the global charges $J_R, J_1-J_2$ and $J_f$ commute with the chosen
supercharges $\CQ_+^{++}$ and $\CQ_+^{--}$. Thus, states with $\bar L_0\not=0$ come in
pairs whose contributions to the partition sum (\ref{def}) vanish. When the Higgs theory is non-trivial (for $F>N$),
one can think of this partition sum as  the (equivariant) elliptic genus of this theory, but it can be defined
when  the infrared SCFT is trivial as well, and we will in fact be interested in this case below. Thus, we will refer to (\ref{def}) as a supersymmetric index.

The index (\ref{def}) is independent of the FI parameter $\xi$, since
the FI term is $\CQ_+$-exact with $\CQ_+=\CQ_+^{++}+\CQ_+^{--}$ and crucially
the asymptotic behavior of the potential
on field space is independent of this parameter.
In particular, the superficially flat potential on the Coulomb
branch at $\xi=0$ is lifted by the nonzero background gauge field that couples to $J_1-J_2$.

As mentioned above, for $\xi\not=0$ the Coulomb branch is lifted, so in that case the index must
come from the Higgs branch. In particular, for the case $F=N$ where the vacuum is isolated
(i.e. all field theoretic degrees of freedom are massive), the index must be equal to $1$,
\begin{align}
  \CE(\t;z,\z_i) = 1\ .
  \label{result}
\end{align}
This index was recently computed in \cite{Benini:2013xpa,Gadde:2013dda}
using localization.
Below we will use their results to verify (\ref{result}).

For $\xi=0$, we saw in the previous section that the infrared limit of the gauge theory with $F=N$
is a direct sum of the $N$ dimensional Coulomb branch, mixed branches in which the rank of the unbroken
subgroup of $U(N)$  is lower than $N$, and the isolated Higgs vacuum. The index is the sum of the contributions of the
various branches. The $\xi$ independence of the index (\ref{def}) implies that the result (\ref{result})
must still be valid when $\xi=0$, and it is natural to ask which branch(es) it is coming from.
We will see later that the Coulomb and mixed branches do not contribute to the index.
Hence, the result (\ref{result}) must come from the conjectured isolated Higgs vacuum.
We conclude that this result requires the existence of the quantum Higgs vacuum.

The above statements must be true on general grounds, but it is useful to verify them using the formalism of \cite{Benini:2013xpa}. It is convenient to rewrite the index (\ref{def}) in the form
\begin{align}
  \CE(\t;\vec z) = \text{Tr}_{\CH_\text{RR}}\left[ (-1)^F q^{L_0} \bar q^{\bar L_0}
  e^{2\pi i \vec z \cdot \vec J} \right]\ ,
\end{align}
where $\vec J = (J_R,J_1-J_2,J_f)$. The authors of \cite{Benini:2013xpa} showed that for
any $\CN=(0,2)$ supersymmetric gauge theory one can write $\CE$ as
\begin{align}
  \CE(\t; \vec z) = \frac{1}{|W|} \sum_{u_\ast \in \mathfrak{M}_\eta} \JKres_{u_\ast}(Q_\ast,\eta)
  Z_\text{1-loop}(u,\vec z) \ ,
  \label{formula1}
\end{align}
where $u$ denotes the holomorphic coordinates parametrizing the moduli space of flat gauge connection
on the torus, $u \in \left(\mathbb{C}/\mathbb{Z}+\t\mathbb{Z} \right)^r$ for a rank $r$
gauge group, and $|W|$ is the order of the Weyl group.
The one-loop determinant $Z_\text{1-loop}$ is
\begin{align}
  Z_\text{1-loop}(u) =  Z_\text{vec}(u) \cdot Z_\text{chi}(u) \cdot Z_\text{fer}(u)
\end{align}
with the contributions from $(0,2)$ vector, chiral and Fermi multiplets given by
\begin{align}
  Z_\text{vec}(u) & = \left[\frac{2\pi \eta(\t)^2}{i}  \right]^r
  \prod_{a:\text{roots}} \frac{i\vartheta_1(\t,\a\cdot u)}{\eta(\t)}\ ,
  \nonumber \\
  Z_\text{chi}(u) & = \prod_{\r:\text{weights}} \frac{i\eta(\t)}{\vartheta_1(\t,\r\cdot u
  + \vec z \cdot \vec J_c)}\ ,
  \nonumber \\
  Z_\text{fer}(u) & = \prod_{\r:\text{weights}}
  \frac{i\vartheta_1(\t,\r\cdot u + \vec z\cdot \vec J_f)}{\eta(\t)}\ ,
  \label{formula2}
\end{align}
where $\vec J_c$ and $\vec J_f$ are the $\vec J$ charges of chiral and Fermi multiplets.
Finally, `JK-Res' is the Jeffrey-Kirwan residue whose definition we review next.

Each $\vartheta_1$ factor in the denominator of $Z_\text{1-loop}$ (\ref{formula2}) defines a hyperplane in
the $u$-plane,
\begin{align}
  \rho\cdot u + \vec z \cdot \vec J_c = 0\ ,
\end{align}
where $Z_\text{1-loop}(u)$ becomes singular. Such a hyperplane is associated with a charge vector
$\rho$ in $\mathbb{R}^r$. Let us denote by $\mathfrak{M}$ the set of singular points $u_\ast$ where $n\geq r$ hyperplanes intersect.
We focus on the case $n=r$, which is the case of interest for the theories discussed below. The Jeffrey-Kirwan residue depends on a choice of an $r$-component vector $\eta$. For given $\eta$, one can define a set $\mathfrak{M}_\eta$ of singular points in $\mathfrak{M}$
where $r$ hyperplanes intersect and the corresponding
$r$ charge vectors generate a positive cone containing the vector $\eta$,
\begin{align}
  \eta \in \text{Cone}[\rho_1,\rho_2,..,\rho_r]\ .
\end{align}
In the case of $n=r$, the Jeffrey-Kirwan residue is defined as follows
\begin{equation}
\def\arraystretch{1.2}
 \JKres_{u_\ast}
 \frac{1}
 {(\rho_1\cdot u)\cdots (\rho_r\cdot u)}
 = \left\{
   \begin{array}{ll}
    \displaystyle
    \frac1{|\text{det}(\rho_1\cdots \rho_r)|}~~ &
    \text{if}~u_\ast \in \mathfrak{M}_\eta \\
    0 & \text{otherwise}.
   \end{array}
   \right.
\def\arraystretch{1}~
\label{JKDef}
\end{equation}
For more details of the Jeffrey-Kirwan residue see
\cite{Benini:2013xpa,Hosomichi:2014rqa}. The index is of course independent of the choice of $\eta$.
For rank one gauge group with $\eta>0$ ($\eta<0$),
the index $\CE(\t,\vec z)$ becomes the sum of residues of $Z_\text{1-loop}(u)$ at the poles
$u=u_\ast$ associated with the fields of positive (negative) charge.

As a warm-up exercise, let us start with a $U(1)$ theory with a single charged
hyper multiplet of charge $+1$. From (\ref{formula2}) one can show that the one-loop
factor corresponding to the $(4,4)$ vector multiplet is
\begin{align}
  Z_{vec}(u) = \frac{i\eta(q)^3}{\vartheta_{1}(\t,\z_1-z)}
  \frac{\vartheta_{1}(\t,2\z_1)}{\vartheta_{1}(\t,\z_1+z)}
\end{align}
while the contribution from the charged hyper multiplet is
\begin{align}
  Z_{hyper}(u) & = \frac{\vartheta_{1}(\t,u-z)}{\vartheta_{1}(\t,u-\z_1)}
  \frac{\vartheta_{1}(\t,-u-z)}{\vartheta_{1}(\t,-u-\z_1)} \ .
\end{align}
Choosing $\eta>0$, the index thus becomes
\begin{align}
  \CE(\t;\xi,z) & = \Res_{u=\z_1} Z_{vec}(u) \cdot Z_{hyper}(u) = 1\ .
  \label{result1}
\end{align}

One can further show that the contribution to the index from the Coulomb branch
vanishes. The low-energy theory on the Coulomb branch can be described by
two $(0,2)$ neutral chiral multiplets $\Sigma,\tilde \Sigma$ and two neutral Fermi multiplets
$\Upsilon,\tilde \Upsilon$. Here the Fermi multiplet $\Upsilon$ contains the field
strength $F_{01}$. The global charges are given by $\vec J_\S=(-1,1)$, $\vec J_{\tilde \S}=(1,1)$,
$\vec J_\Upsilon = (0,0)$ and $\vec J_{\tilde \Upsilon}=(2,0)$. Since the Fermi multiplet $\Upsilon$
is neutral under both $J_R$ and $J_1 - J_2$, one can show from (\ref{formula2}) that
the contribution from the Coulomb branch to the index vanishes. This is essentially because
the fermionic zero mode in $\Upsilon$ will pair states of opposite fermion number and equal charges $\vec J$.

We next turn to the case of general $N$, i.e.,
$\CN=(4,4)$ $U(N)$ gauge theory with $F=N$ fundamental hyper multiplets.
We use the standard unit vectors $e_i$ in $\mathbb{R}^N$, $i=1,2,..,N$, to parameterize the weight vectors of the fundamental representation. The root
vectors can be expressed as $e_i-e_j$ with $i\neq j$. The one-loop determinant is in this case
\begin{align}
  Z_{vec}(u) & = \left[ \frac{i\eta(q)^3}{\vartheta_{1}(\t,\z_1-z)} \frac{\vartheta_{1}(\t,2\z_1)}
  {\vartheta_{1}(\t,\z_1+z)} \right]^N
  \nonumber \\ & \times
  \prod_{i\neq j} \frac{\vartheta_{1}(\t,(e_i-e_j) \cdot u )}
  {\vartheta_{1}(\t,(e_i-e_j)\cdot u + \z_1-z)}
  \frac{\vartheta_{1}(\t,(e_i-e_j)\cdot u + 2 \z_1)}{\vartheta_{1}(\t,(e_i-e_j)\cdot u + \z_1+z)}\ ,
  \label{vec}
\end{align}
and
\begin{align}
  Z_{hyper}(u) = \prod_{i} \prod_{\m : \text{weights} }
  \frac{\vartheta_{1}(\t,e_i\cdot u + \m \cdot \z_2 -z)}{\vartheta_{1}(\t,e_i\cdot u + \m\cdot \z_2  - \z_1 )}
  \frac{\vartheta_{1}(\t,-e_i\cdot u - \m \cdot \z_2 -z)}{\vartheta_{1}(\t,-e_i\cdot u- \m\cdot \z_2  - \z_1 )}\ ,
  \label{hyper}
\end{align}
where $\m$ denote the weight vectors of the fundamental representations under the flavor group $SU(N)$.

To compute the index\footnote{Our analysis is parallel to that of \cite{Benini:2013xpa}.},
we first choose the $N$-component vector $\eta$ as follows
\begin{align}
  \eta = e_1 + e_2 + .. + e_N\ .
\end{align}
One can show that any set of charge vectors generating a positive cone that contains the vector $
\eta$ should have either a pair of weight vector $e_i$ and root vector $e_j-e_i$ for certain $i$ and $j$
or weight vectors only.

In the case that the set of charge vectors have a pair of $e_i$ and $e_j-e_i$, the Jeffrey-Kirwan
residue becomes trivial. This is because the corresponding singular points satisfy
\begin{align}
  \left( e_j - e_i \right)\cdot u_\ast + \z_1 \pm z & = 0
  \nonumber \\
  e_i \cdot u_\ast + \m \cdot \z_2 - \z_1 & = 0
\end{align}
with a weight vector $\m$ in the fundamental representation of $SU(N)$.  This implies that
\begin{align}
  e_j \cdot u_\ast + \mu \cdot \z_2 \pm z = 0
\end{align}
for which one of the factors in the numerator of $Z_{hyper}$ in (\ref{hyper}) vanishes.

For the cone generated by the weight vectors $\{e_1,e_2,..,e_N\}$, there are corresponding $N^N$ singular
points satisfying
\begin{align}
  e_1 \cdot u_\ast + \mu_1 \cdot \z_2 - \z_1 & = 0\ ,
  \nonumber \\
  e_2 \cdot u_\ast + \mu_2 \cdot \z_2 - \z_1 & = 0\ ,
  \nonumber \\ & \vdots
  \nonumber \\
  e_N \cdot u_\ast + \mu_N \cdot \z_2 - \z_1 & = 0\ ,
\end{align}
for arbitrary $N$ weight vectors $\mu_i$. However if two such weight vectors coincide,
for instance $\mu_1=\mu_2$, the Jeffrey-Kirwan residue vanishes because
\begin{align}
  \left( e_1- e_2\right)\cdot u_\ast = 0
\end{align}
at which the factor $\vartheta_{1}(\t,(e_1-e_2)\cdot u)$ in the numerator of $Z_{vec}$ in (\ref{vec}) becomes zero.
Thus we have $N!$ singular points with $\mu_i\neq \mu_j$ for any $i,j$, at each of which the
Jeffrey-Kirwan residue can be shown to give one. Dividing by the order of the Weyl group (\ref{formula1}) one finds
\begin{align}
  \CE(\t;z,\xi_i) = 1\ .
\end{align}

As in the $U(1)$ case, one can verify that the contributions to the index from the Coulomb and
mixed branches vanish due to the existence of the neutral Fermi multiplet. Therefore, one concludes that
there must exist an isolated quantum Higgs vacuum in the
$\CN=(4,4)$ $U(N)$ theory coupled to $N$ flavors.

\section*{Acknowledgments}

We thank Piljin Yi for helpful discussions.
JH acknowledges the support of NSF grant 1214409.
DK is supported in part by DOE grant DE-FG02-13ER41958.
The work of SL is supported in part by the Ernest Rutherford fellowship of
the Science and Technology Facilities Council ST/J003549/1.

\bibliographystyle{JHEP}
\bibliography{QH}

\end{document}